\documentclass[runningheads]{llncs}
\usepackage{amsfonts}
\usepackage{graphicx}
\usepackage{diagbox}
\usepackage[T1]{fontenc}
\usepackage{color, colortbl}
\definecolor{teagreen}{rgb}{0.9, 0.97, 0.90}
\definecolor{wildblueyonder}{rgb}{0.64, 0.68, 0.82}
\usepackage[hyphens]{url}
\usepackage{caption}
\usepackage{subcaption}
\usepackage{color,soul}
\usepackage{amsmath}
\DeclareMathSymbol{\Beta}{\mathalpha}{operators}{"42}

\begin{document}
\title{HMAR: Hierarchical Masked Attention for Multi-Behaviour Recommendation}
\titlerunning{HMAR for Multi-Behaviour Recommendation}
%

%



\author{Shereen Elsayed \inst{1} \and
Ahmed Rashed\inst{2} \and
Lars Schmidt-Thieme\inst{1}}
\authorrunning{S. Elsayed et al.}
%
\institute{Information Systems and Machine Learning Lab (ISMLL)
\& VWFS Data Analytics Research Center (VWFS DARC), University of Hildesheim
, Hildesheim, Germany  \ \email{\{elsayed,schmidt-thieme\}@ismll.uni-hildesheim.de} \and
Data Analytics \& AI Engineering
Volkswagen Financial Services AG, Braunschweig, Germany \
\email{{ahmed.rashed}@vwfs.io}}
%
%

%

%

%

%
\maketitle            
\begin{abstract}

In the context of recommendation systems, addressing multi-behavioral user interactions has become vital for understanding the evolving user behavior. Recent models utilize techniques like graph neural networks and attention mechanisms for modeling diverse behaviors, but capturing sequential patterns in historical interactions remains challenging. To tackle this, we introduce Hierarchical Masked Attention for multi-behavior recommendation (HMAR). Specifically, our approach applies masked self-attention to items of the same behavior, followed by self-attention across all behaviors. Additionally, we propose historical behavior indicators to encode the historical frequency of each item's behavior in the input sequence. Furthermore, the HMAR model operates in a multi-task setting, allowing it to learn item behaviors and their associated ranking scores concurrently. Extensive experimental results on four real-world datasets demonstrate that our proposed model outperforms state-of-the-art methods. Our code and datasets are available here \footnote{\url{https://github.com/Shereen-Elsayed/HMAR}}. 
\end{abstract}
\section{Introduction}
Multi-behavioural recommendation aims to integrate all user behaviors into the next-item recommendation task. Such kind of approaches leverage all user behaviors, yielding more insightful recommendations. For example, in retail platforms, clicking on an item and adding it to favorites can indicate an intent to purchase.

Recommender systems approaches have evolved over time, from traditional matrix factorization to modern neural network-based methods like Neural Collaborative Filtering (NCF) \cite{he2017neural}, SASRec \cite{kang2018self} and BERT4Rec \cite{sun2019bert4rec} for sequential recommendations, CARCA \cite{rashed2022context}, SSE-PT \cite{zhou2020s3} and TiSASRec \cite{li2020time} for context and attribute-aware recommendations. However, these techniques primarily rely only on the user's purchase interactions within retail platforms. Recently, newly developed models aim to harness additional user-item interaction data, particularly in multi-behavior scenarios. Graph-based approaches like MB-GCN \cite{jin2020multi} and MB-GMN \cite{xia2021graph} construct user-item and item-item graphs to model diverse behaviors. More recent methods, like MB-STR \cite{yuan2022multi} and MBHT \cite{yang2022multi}, employ attention mechanisms. Yet, they often struggle to capture comprehensive behavioral representations and capture complex sequential patterns in the user history.

To bridge this gap, we introduce HMAR, a hierarchical masked attention model that captures latent behavioral representations while preserving temporal sequences in the data. Our contributions can be summarized as follows:
\begin{itemize} 
\item We present the HMAR model for the multi-behavior recommendation, incorporating hierarchical masked attention to learn comprehensive behavioral representations. Furthermore, we employ multi-task learning to understand both user intentions and interaction types.

\item \sloppy Within the HMAR model, we introduce hierarchical masked attention that models deep behavioral relationships within and across behaviors while preserving input sequence dynamics. We also include historical behavior indicators (HBI) for tracking item behavior frequency.
\item  Extensive experiments on four real-world datasets show that our proposed model HMAR outperforms previous state-of-the-art multi-behavioral recommendation models.

\end{itemize}



\begin{figure}[ht]
  \centering
  \includegraphics[scale=0.8]{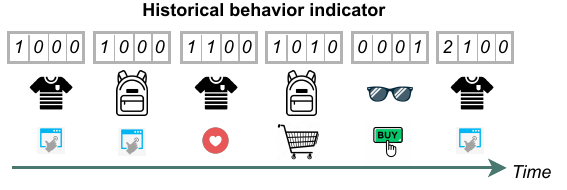}     
  \caption{ Historical behavior indicator example }
  \label{fig:3}
\end{figure}
\section{Methodology}
In this section, we present the HMAR model, employing masked attention historical behavior indicator features to capture complex multi-behavioral sequential representations. Additionally, we present the multi-task learning aspect, where HMAR learns both item scores and behaviors through an auxiliary classification task.

\makeatletter
\DeclareRobustCommand*\cal{\@fontswitch\relax\mathcal}
\makeatother
\subsection{Problem Formulation}
In multi-behavior settings, given a set of items ${\cal I}:=\{1,\ldots,$ $I\}$ and a set of users ${\cal U}:=\{1,\ldots,$ $U\}$, each user $u$ has a historical ordered sequence $S$ of item interactions ${S^{u}}:= \{v_{1}^{u},\ldots, v_{|S_t^u|}^{u}\}$. Given $K$ behaviors, each item in the sequence has a corresponding behavior type $b \in { \cal B}:=\{1,\ldots,$ $K\}$, such as buy and add to favorite.  Our goal is to predict the next item to be interacted with based on the primary behavior, while other behaviors serve as auxiliary information to better model the user behavior over time.

\begin{figure*}[ht]
  \centering
  \includegraphics[scale=0.36]{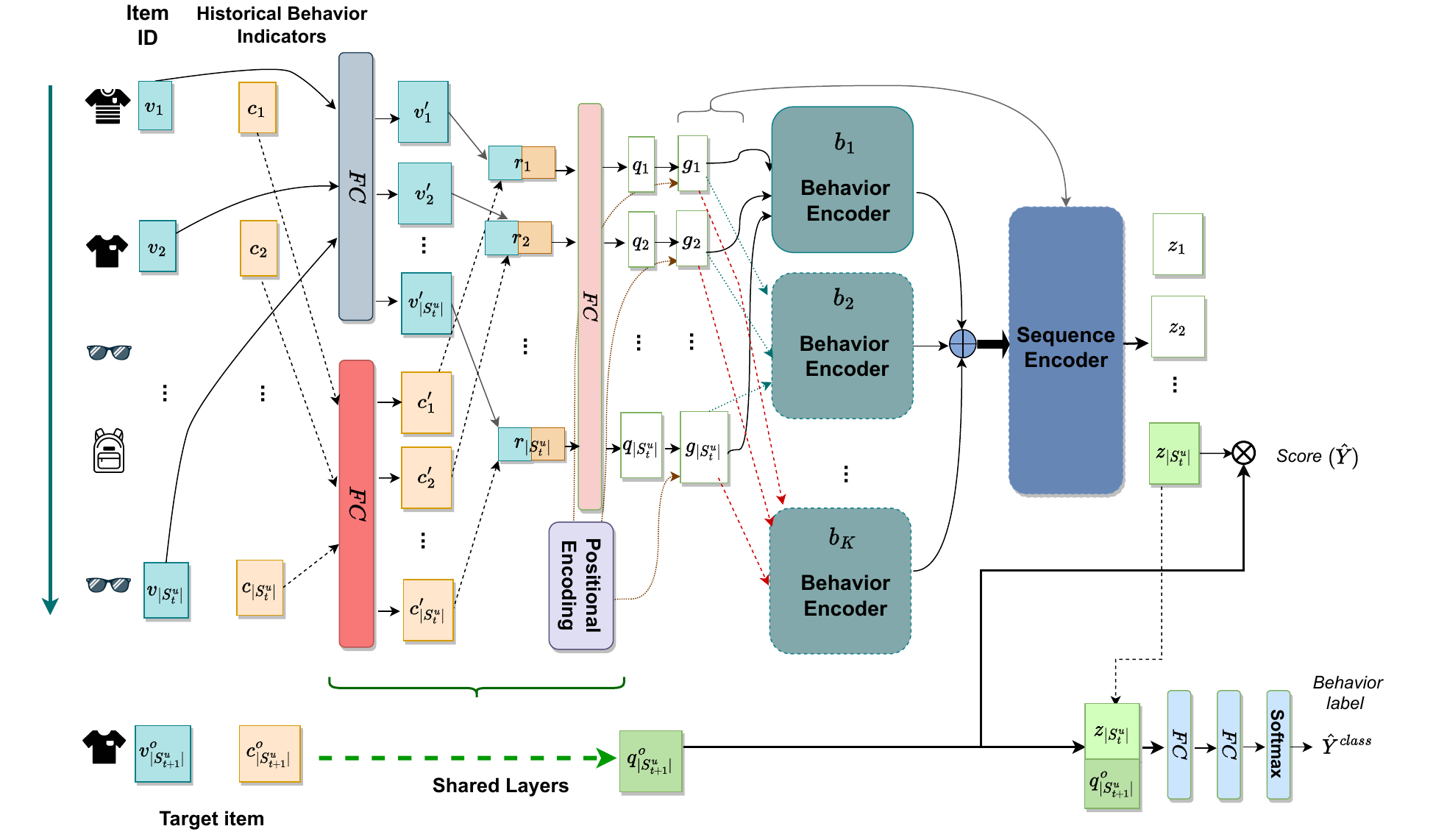}     
  \caption{Hierarchical Masked Attention Model Architecture}
  \label{fig:1}
\end{figure*}

\subsection{HMAR}
Figure \ref{fig:1} illustrates the model architecture. It consists of Items and Behaviors Encoding, Hierarchical masked attention, and Multi-task learning.

\subsubsection{\textbf{Items and Behaviors Encoding}}

As discussed earlier, each user has an input sequence of historical item interactions $S^u$ and a corresponding behaviors sequence $B^u=[ b_1, b_2, ..., b_{|S^u|}]$. To construct the latent item representation, We first embed each item in the input sequence using a fully connected layer to obtain the initial embeddings as: 

\begin{equation}
    v'_{j}= v_j W_v + \Beta_v
    \label{eq:1}
\end{equation} 
\noindent where $W_v \in \mathbb{R}^{I \times d}$ is the weight matrix, $I$ is the number of items, and $d$ is the items embedding dimension and $\Beta_v$ is the bias term.
To encode item behaviors, we use historical behavior indicators $c_j$ that track the frequency of each behavior type on an item throughout the sequence. Figure \ref{fig:3} shows an example of the historical behavior indicators in a sequence. We also embed these indicators using a fully connected layer. This allows the model to learn the order and the frequency of user behaviors on items. 
\begin{equation}
    c'_{j}= c_j W_c +\Beta_c
    \label{eq:2}
\end{equation}
\noindent where $W_c \in \mathbb{R}^{N \times d}$ is the weight matrix, $N$ is the number of different historical behavior indicators, $d$ is the context embedding dimension and $\Beta_c$ is the bias term. For simplicity, we use the \textbf{same dimension $d$} for all embedding layers. 

Afterward, we concatenate both embeddings to get the combined item encoding $r_j$. Then, feed it to another fully connected layer for embedding item and historical behavior indicator together:

\begin{align} 
    r_{j}= \text{concat}_{col} \left( v'_j, c'_j  \right),\ \ \ \ \ \ q_{j}= r_j W_r +\Beta_r
\end{align}


where $W_r \in \mathbb{R}^{2d \times d}$ is the weight matrix, $d$ is the layer embedding dimension, and $\Beta_r$ is the bias term. 
Finally, a learnable positional encoding is added to the final embedding to indicate the  position in the input interactions sequence:
\begin{equation}
    g_{j}= q_{j} +P_j
    \label{eq:5}
\end{equation}

\begin{figure}[ht]
  \centering
  \includegraphics[scale=0.40]{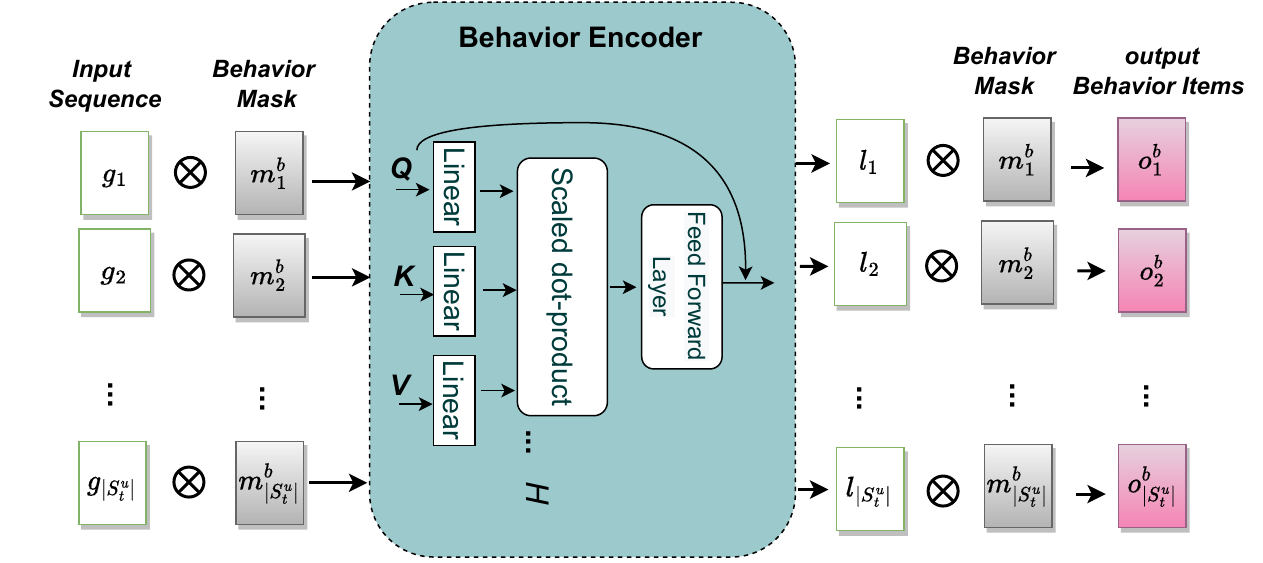}     
  \caption{ Behavior Encoder Architecture}
  \label{fig:2}
\end{figure}

\subsubsection{\textbf{Hierarchical Masked Attention}}
In sequential multi-behavior recommendation, sequence order is crucial for an accurate next-item recommendation. In our work, we model behavioral interactions using two-stage self-attention to preserve input sequence order. The first stage is the \textbf{Behavior Encoder}, which encodes the items with the same behavior in the sequence, followed by a \textbf{Sequence Encoder}, which encodes the whole sequence and allows the modeling of different behaviors in the input sequence.

\subsubsection{Behavior Encoder}
Recent approaches use graphs to gather multi-behavior information, potentially sacrificing the input sequence's sequential pattern. To maintain the items' order while obtaining better representations, we employ a behavior mask $M_{b}^{u}$, created from the sequence's items, to filter out items not associated with a specific behavior $b$. Accordingly, we multiply the item sequence embedding ${G^{u}}:= [g_{1}, g_{2}, ..., g_{|S^{u}_{t}|}]$ and with the corresponding mask of the specified behavior $M_{b}^{u}:= [m^b_1, m^b_2, ..., m^b_{|S^{u}_{t}|}]$ as shown in Figure \ref{fig:2}.

This yields a number of behavior encoder blocks equivalent to the number of behaviors $K$. The masked latent embeddings are then fed into a fully connected layer to get ${E_{b}^{u}}:= \{e_{1}^{b},\ldots, e_{|S_t^u|}^{b}\}$ followed by a multi-head self-attention component to sequentially encode items with the same behavior as follows:
\begin{equation}
     e_{j}^{ b} = \left(g_{j} * m^b_{j} \right) W_{e} + \Beta{e},\ \ W_{e} \in \mathbb{R}^{d \times d},\ \ \Beta_{e} \in \mathbb{R}^{d} 
\end{equation}


\begin{equation} 
X_{b}^{u}=\textrm{SA}(E_{b}^{u})=  
\text{concat}_{col} \left( \textrm{Att}(E_{b}^{u}\textbf{W}_{h}^{{Q}^b}, E_{b}^{u}\textbf{W}_{h}^{{K}^b}, E_{b}^{u}\textbf{W}_{h}^{{V}^b}) \right)_{h=1:H}
 \label{eq:8}
\end{equation} 
\noindent where $\textbf{W}^{Q^b}_{h}$, $\textbf{W}^{K^b}_{h}$, $\textbf{W}^{V^b}_{h} \in \mathbb{R}^{d \times \frac{d}{H}}$ represent the linear projection matrices of the head at index $h$, and $H$ is the number of heads.
$X_{b}^{u}$ represents the column-wise concatenation of the attention heads. Finally, we have the point-wise feed-forward layers to obtain the component's final output representations $F_{b}^{u} \in \mathbb{R}^{ |{S^{u}_{t}}| \times d }$, $F_{b}^{u} = \textrm{FFN}(X_{b}^{u}) $ as follows:
\begin{equation} 
\textrm{FFN}(X_{b}^{u}) =
\text{concat}_{row} \left( ReLU ({X_{b,j}^{u}} \textbf{W}^{(1)^b} + \Beta^{(1)^b})\textbf{W}^{(2)^b} + \Beta^{(2)^b}  \right)_{j=1:|{S^{u}_{t}}|}
 \label{eq:9}
\end{equation}
\noindent where $\textbf{W}^{(1)^b}$, $\textbf{W}^{(2)^b} \in \mathbb{R}^{d \times d}$ are the weight matrices of the two feed-forward layers, and $\Beta^{(1)^b}$, $\Beta^{(2)^b} \in \mathbb{R}^{d}$ are their bias vectors. The initial embedding sequence is added via a ReZero \cite{bachlechner2021rezero} residual connection to the behavior encoder output to obtain as:
\begin{equation} L_{b}^{u} = F_{b}^{u} + \gamma  \left(G^u * M_{b}^{u} \right)
\label{eq:9}
\end{equation}
\noindent where $\gamma$ is a learnable weight initialized with zero to adjust the contribution of $\left(G^u * M_{b}^{u} \right)$. The output is multiplied by the behavior mask again to mask the positions of the other behaviors as follows: 
\begin{equation}
    O_{b}^{u}= L_{b}^{u}  * M_{b}^{u}
    \label{eq:10}
\end{equation}

Finally, we combine the output sequences of each behavior \sloppy ${O_{b}^{u}}:= [o^{b}_{1}, o^{b}_{2}, ..., o^{b}_{|S^{u}_{t}|}]$ by summing them element-wise to generate the first stage multi-behavioral latent sequence embeddings $\mathcal{O}^{u}$ as follows:
\begin{equation}
      \mathcal{O}^{u}=  \sum^{K}_{b=0} O_{b}^{u}
      \label{eq:11}
\end{equation}

\subsubsection{\textbf{Sequence Encoder}}
In the second phase of hierarchical attention, we utilize a multi-head self-attention component across all behaviors while maintaining the original temporal order of the sequence. Specifically, once we've constructed the sequence comprising all behaviors, denoted as $\mathcal{O}^{u}$, the sequence encoder employs attention mechanisms over all items within the sequence. This enables the model to capture dependencies among various behaviors within the sequence, yielding an intra and inter-multi-behavioral latent representation for each item:
\begin{equation} 
A^{u}=\textrm{SA}(\mathcal{O}^{u})= \\
\text{concat}_{col} \left( \textrm{Att}(\mathcal{O}^{u}\textbf{W}^{{Q}}_{h}, \mathcal{O}^{u}\textbf{W}^{{K}}_{h}, \mathcal{O}^{u}\textbf{W}^{{V}}_{h}) \right)_{h=1:H}
 \label{eq:13}
\end{equation}

\noindent where $\textbf{W}^{Q}_{h}$, $\textbf{W}^{K}_{h}$, $\textbf{W}^{V}_{h} \in \mathbb{R}^{d \times \frac{d}{H}}$ represent the linear projection matrices of the head at index $h$, and $H$ is the number of heads.
$A^{u}$ represents the column-wise concatenation of the attention heads.
Additionally,for the model stability, we add a residual connection between sequence attention output $A^{u}$ and sequence items $G^{u}$:
\begin{equation}
   {A^{u}}'= A^{u} + G^{u} 
   \label{eq:14}
\end{equation}
Finally, we have the point-wise feed-forward layers to obtain the component's output representations $Z^{u} \in \mathbb{R}^{  |{S^{u}_{t}}| \times d } $ as follows:

\begin{equation} 
Z^{u} = \textrm{FFN}({A^{u}}') = 
\text{concat}_{row} \left( ReLU ({A^{u}}' \textbf{W}^{(1)^Z} + \Beta^{(1)^Z})\textbf{W}^{(2)^Z} + \Beta^{(2)^Z}  \right)_{j=1:|{S^{u}_{t}}|}
 \label{eq:15}
\end{equation}

\noindent where $\textbf{W}^{(1)^Z}$, $\textbf{W}^{(2)^Z} \in \mathbb{R}^{d \times d}$ are the weight matrices of the two feed-forward layers, and $\Beta^{(1)^Z}$, $\Beta^{(2)^Z} \in \mathbb{R}^{d}$ are their bias vectors.

\subsection{Multi-Task Learning}
In order to train the model, we utilize a multi-task learning approach to learn item ranking and behavior type simultaneously. For efficient training, we adopt the SASRec \cite{kang2018self} process, which uses autoregressive training, considering only historical items for each target item recommendation. The input sequence, $S^u$, has a fixed length, achieved by padding or truncating. The target sequence is constructed by right-shifting the input sequence and including the last item interaction. On the other hand, negative items are sampled using random items not in the user's interaction history $v \notin S^u$.

For item ranking, we calculate the final score by taking the dot product of the last item embedding $z_{|S^{u}_{t}|}$ from the sequence encoder and the target item embedding $q^o_{|S^{u}_{t+1}|}$ as follows:
\begin{equation}
    \hat{Y}_{t+1} = \sigma(z_{|S^{u}_{t}|} \cdot q^o_{|S^{u}_{t+1}|})
    \label{eq:16}
\end{equation}
\noindent Where $\sigma$ is a sigmoid function. 
In multi-behavior datasets, interactions have varying importance for next-item recommendation. To handle this, we introduce weighting factors $\alpha_b$ in the loss function, one for each behavior. The number and values of $\alpha_b$ may differ based on the dataset's behavior count and their importance. Thus, our weighted binary cross-entropy objective for multi-behavior recommendation is as follows:

\begin{equation}
      \mathcal{L}_{rank}=  - \sum_{S^u \in \mathcal{S}} \sum_{t =0} ^{|S^u|} [ \alpha_b log{(\hat{Y}^{O^{(+)}}_{t})} + \beta (log{(1-\hat{Y}^{O^{(-)}}_{t})})]
       \label{eq:17}
\end{equation}
\noindent where $\hat{Y}^{O^{(+)}}_{t}$ are the output scores for the positive samples and $\hat{Y}^{O^{(-)}}_{t}$ are the output scores for the negative samples, $\mathcal{S}$ is the set of all sequences, $\alpha_b$ is the behaviors weights, and lastly, $\beta$ is the sampling weight. 

In this work, we also propose learning target interaction behavior type as an auxiliary task enhancing the model's ability to understand user behavior patterns and anticipate the behavior type of the next interaction. The behavior-type task involves a two-layer neural network that takes concatenated item embeddings $z_{|S^{u}_{t}|}$ and $q^o_{|S^{u}_{t+1}|}$ as input, as follows:
\begin{equation} 
\hat{Y}^{class} = \text{Softmax}((\text{concat}_{col}(z_{|S^{u}_{t}|}, q^o_{|S^{u}_{t+1}|}) \textbf{W}^{(1)} + \Beta^{(1)})\textbf{W}^{(2)} + \Beta^{(2)})
 \label{eq:18}
\end{equation}
\noindent where $\textbf{W}^{(1)}\in \mathbb{R}^{2d \times d}$ and  $\textbf{W}^{(2)} \in \mathbb{R}^{d \times d}$ are the weight matrices of the two fully connected layers, and $\Beta^{(1)}, \Beta^{(2)} \in \mathbb{R}^{d}$ are their bias vectors. 
Then, we apply softmax function on the output to get the labels scores of each behavior and utilize the cross entropy classification loss as follows:

\begin{equation}
      \mathcal{L}_{class}=  - \sum_{S^u \in \mathcal{S}} \sum^{|S^u|}_{t=0} \sum^K_{j=0} [ Y_{t_j}^{class}log{(\hat{Y}^{class}_{t_j})} ]
       \label{eq:20}
\end{equation}

\noindent where $Y_{t_j}$ is the true behavior labels and $\hat{Y}_{t_j}^{class}$ is the predicted classification scores. 
Finally, the model is optimized using ADAM optimizer \cite{kingma2014adam}. The final model loss can be defined as a weighted sum of the two losses:
\begin{equation}
      \mathcal{L}_{model}=  \mathcal{L}_{rank} + \theta \mathcal{L}_{class}
       \label{eq:21}
\end{equation}
\noindent where $\theta$ is a hyperparameter controlling the contribution of the classification task to the final loss.

\begin{table*}
\centering
\scalebox{0.8}{
\begin{tabular}{c|c|c|c|c}
\hline
Dataset     & Interactions & User\# & Item\# & Behaviors \\
\hline
Taobao       & 7,658,926 & 147,894 & 99,037 & Page View, Fav., Cart, Buy \\
Yelp         & 1,400,000 & 19,800 & 22,734 & Tip, Dislike, Neutral, Like \\
MovieLens    & 9,922,036 & 677,88 & 8704 & Dislike, Neutral, Like \\
Tianchi      &  4,619,389 & 25,000 & 500,900 & Page View, Fav., Cart, Buy   \\ 
\hline
\end{tabular}
}
\caption{Datasets Statistics}
\label{tab:1}

\end{table*}

\begin{table*}[!]
\setlength{\tabcolsep}{4pt}
\scalebox{0.9}{
\centering
\begin{tabular}{c|cc|cc}
\hline

Method    & \multicolumn{2}{c}{Taobao} & \multicolumn{2}{c}{Yelp} \\
            & HR@10 & NDCG@10 & HR@10 & NDCG@10 \\
\hline  
\multicolumn{5}{c}{\textit{Sequential and Context-Aware Recommendation Methods}}\\
\hline  
SASRec \cite{kang2018self} & 0.390 $\pm$ \tiny{$4.2E{-3}$} & 0.249 $\pm$ \tiny{$8E{-4}$}& 0.853  $\pm$ \tiny{$1.7E{-3}$}   &  0.5601 $\pm$ \tiny{$5.1E{-3}$}  \\

SSE-PT \cite{wu2020sse} & 0.393 $\pm$ \tiny{$2.7E{-3}$} & 0.232 $\pm$ \tiny{$3E{-4}$} & 0.857 $\pm$ \tiny{$1.5E{-2}$} &  0.572 $\pm$ \tiny{$1.8E{-2}$}  \\
CARCA \cite{rashed2022context} & \underline{0.769 $\pm$ \tiny{$4E{-3}$}} &   \underline{0.662 $\pm$ \tiny{$5E{-3}$} } & 0.854 $\pm$ \tiny{$2E{-3}$}  & 0.589 $\pm$ \tiny{$4E{-3}$} \\
\hline
\multicolumn{5}{c}{\textit{Multi-Behavior Recommendation Methods}}\\
\hline 
MATN \cite{xia2020multiplex}& 0.354$\dagger$ & 0.209$\dagger$  & 0.826$\dagger$  & 0.530$\dagger$  \\
MB-GCN \cite{jin2020multi}& 0.369$\dagger$ & 0.222$\dagger$  &  0.796$\dagger$ & 0.502$\dagger$  \\
MB-GMN \cite{xia2021graph} & 0.491$\dagger$ & 0.300$\dagger$  & 0.861  $\pm$ \tiny{$8E{-3}$}  & 0.570  $\pm$ \tiny{$1.2E{-2}$} \\  
KHGT \cite{xia2021knowledge}& 0.464$\dagger$ & 0.278$\dagger$  & 0.880$\dagger$  & 0.603$\dagger$ \\
KMCLR \cite{xuan2023knowledge}  & 0.4557 $\dagger$  & 0.2735 $\dagger$  & \underline{0.8897} $\dagger$  & 0.6038 $\dagger$  \\
MB-STR \cite{yuan2022multi} & 0.768$\dagger$ & 0.608$\dagger$  & 0.882$\dagger$  & \underline{0.624}$\dagger$ \\
MBHT \cite{yang2022multi}& 0.745 $\pm$ \tiny{$6.1E{-3}$}& 0.559 $\pm$ \tiny{$8.4E{-3}$} & 0.885 $\pm$ \tiny{$5E{-4}$} & 0.618 $\pm$ \tiny{$1.4E{-3}$} \\  
\textbf{HMAR (ours)} &  \textbf{0.8515 $\pm$ \tiny{$ 1.8E{-3}$}} & \textbf{0.7294 $\pm$ \tiny{$1.2E{-3}$}} & \textbf{0.9015 $\pm$ \tiny{$5.7E{-5}$}}  & \textbf{0.6374 $\pm$ \tiny{$8.08E{-4}$}} \\
\hline  
Improv.(\%)& 10.663\% & 10.120\%  & 1.349\%  & 2.083\% \\
\hline 
\end{tabular}
}
\caption{Model performance and comparison against baselines on Taobao and Yelp datasets. Published results are indicated by the $\dagger$ symbol.}
\label{tab:2}
\end{table*}


\section{Experiments}
In this section, we aim to address the following research questions: \\
\begin{itemize}
\item \textbf{RQ1:} How does the HMAR model's performance compare to that of state-of-the-art multi-behavior recommendation models?
\item \textbf{RQ2:} What impact do auxiliary behaviors have on the model's performance?
\item \textbf{RQ3:}  What influence does each individual model component have on the overall model performance?

\end{itemize}
\subsection{Experimental Settings}
\subsubsection{Datasets}
We assess our model on four multi-behavioral datasets:\\
\textbf{Taobao \footnote{https://github.com/akaxlh/MB-GMN/tree/main/Datasets/Tmall} \cite{xia2021knowledge,yuan2022multi} :}  Data from the popular Chinese online shopping platform Taobao with behaviors like buy, add-to-cart, add-to-favorite, and pageview. We consider buy behavior as the target behavior. \textbf{Tianchi \footnote{https://tianchi.aliyun.com/competition/entrance/231576/information} :} Collected from Tmall.com, similar to Taobao, it includes buy, add-to-cart, add-to-favorite, and pageview behaviors. \textbf{Yelp \footnote{https://github.com/akaxlh/KHGT/tree/master/Datasets/Yelp} \cite{xia2021knowledge,yuan2022multi} :} Data gathered from the Yelp challenge, categorizing behaviors based on user ratings. Behaviors are split into dislike (rating $\leq$ 2), natural (2 < rating < 4), and like (rating $\geq$ 4). Users can also write venue tips. \textbf{MovieLens \footnote{https://github.com/akaxlh/KHGT/tree/master/Datasets/MultiInt-ML10M} \cite{xia2021knowledge} :} Similar to Yelp, but with movie ratings, categorizing behaviors as dislike, neutral, and like. The main behavior considered for both MovieLens and Yelp datasets is the "like" behavior. Datasets statistics are summarized in Table \ref{tab:1}.

\begin{table*}[!]
\setlength{\tabcolsep}{4pt}
\scalebox{0.9}{
\centering
\begin{tabular}{c|cc|cc}
\hline

Method    & \multicolumn{2}{c}{MovieLens}  & \multicolumn{2}{c}{Tianchi}\\
            & HR@10 & NDCG@10 & HR@10 & NDCG@10\\
\hline  
\multicolumn{5}{c}{\textit{Sequential and Context-Aware Recommendation Methods}}\\
\hline  
SASRec \cite{kang2018self} &  0.911 $\pm$ \tiny{$1E{-3}$}& 0.668 $\pm$ \tiny{$5.1E{-3}$}& 0.659 $\pm$ \tiny{$3E{-3}$}& 0.495 $\pm$ \tiny{$2E{-3}$}  \\

SSE-PT \cite{wu2020sse}  & 0.911 $\pm$ \tiny{$7.1E{-3}$} & 0.657 $\pm$ \tiny{$4.5E{-3}$} & 0.663 $\pm$ \tiny{$1.2E{-2}$} & 0.468 $\pm$ \tiny{$1.3E{-2}$}  \\
CARCA \cite{rashed2022context} & 0.906 $\pm$ \tiny{$2E{-3}$} & 0.665 $\pm$ \tiny{$1E{-3}$} &   0.713 $\pm$ \tiny{$4E{-4}$}&  0.500 $\pm$ \tiny{$1E{-3}$}\\
\hline
\multicolumn{5}{c}{\textit{Multi-Behavior Recommendation Methods}}\\
\hline 
MATN \cite{xia2020multiplex}& 0.847$\dagger$ & 0.569$\dagger$ & 0.714 $\pm$ \tiny{$7E{-4}$}  & 0.485 $\pm$ \tiny{$2E{-3}$} \\
MB-GCN \cite{jin2020multi}&  0.826$\dagger$ & 0.553$\dagger$ & -  & - \\
MB-GMN \cite{xia2021graph} &   0.820  $\pm$ \tiny{$1.1E{-3}$}& 0.530  $\pm$ \tiny{$9E{-4}$} & \underline{0.737  $\pm$ \tiny{$4.3E{-3}$}}  & 0.502  $\pm$ \tiny{$1.9E{-3}$} \\  
KHGT \cite{xia2021knowledge}& 0.861$\dagger$ & 0.597$\dagger$ & 0.652 $\pm$ \tiny{$1E{-4}$}  & 0.443 $\pm$ \tiny{$1E{-4}$} \\
MBHT \cite{yang2022multi} &  \underline{0.913 $\pm$ \tiny{$5.9E{-3}$}} & \underline{0.695 $\pm$ \tiny{$7E{-3}$}} & 0.725 $\pm$ \tiny{$6.3E{-3}$}  & \underline{0.554 $\pm$ \tiny{$4.8E{-3}$}} \\  
\textbf{HMAR (ours)} &  \textbf{0.9412 $\pm$ \tiny{$5.5E{-4}$}} & \textbf{0.7370 $\pm$ \tiny{$1.7E{-3}$}} & \textbf{0.7842 $\pm$ \tiny{$ 1.0E{-3}$}}  & \textbf{0.5974 $\pm$ \tiny{$ 1.7E{-3}$}} \\
\hline  
Improv.(\%)& 3.066\% & 6.043\% & 6.377\%  & 7.761\% \\
\hline 
\end{tabular}
}
\caption{Model performance and comparison against baselines on MovieLens and Tianchi datasets. Published results are indicated by the $\dagger$ symbol.}
\label{tab:3}
\end{table*}
\subsection{Evaluation Protocol}
For a fair comparison, we follow the same evaluation process as recent state-of-the-art methods \cite{xia2021knowledge}\cite{yuan2022multi}. We use a leave-one-out mechanism, training and validating the model with the entire sequence except the last interaction, which is used for testing. We generate 99 negative items for each positive item and assess model performance with Hit Ratio (HR@N) and Normalized Discounted Cumulative Gain (NDCG@N). Higher HR and NDCG values indicate better performance. We report the mean and standard deviation of results from three separate runs to ensure statistical robustness.

\subsubsection{Baselines}
We compare our proposed method against various sequential and multi-behavioral recommendation methods.

\paragraph{\textbf{Sequential and Context-Aware Recommendation Methods.}}
\begin{itemize}
\item \textbf{SASRec} \cite{kang2018self}: A model that utilizes multi-head self-attention to capture the sequential pattern in the users' history, then applies dot product for calculating the items scores.  
\item \textbf{SSE-PT} \cite{wu2020sse}: A state-of-the-art model which incorporates the user embedding into a personalized transformer with stochastic shared embedding regularization for handling extremely long sequences.  
\item \textbf{CARCA} \cite{rashed2022context}: A context-aware sequential recommendation approach employing cross-attention between user profiles and items for score prediction.
\end{itemize}
\paragraph {\textbf{Multi-Behavior Recommendation Models.}}
\begin{itemize}
\item \textbf{MATN} \cite{xia2020multiplex}: A memory-augmented transformer network that utilizes a transformer-based encoder that jointly models behavioral dependencies.

\item \textbf{MB-GCN} \cite{jin2020multi}: A model that represents the data as a unified graph, then employs a graph-convolutional network to learn the node's representation.  

\item \textbf{MB-GMN} \cite{xia2021graph}: Learns the multi-behavior dependencies through graph meta-network.

\item \textbf{KHGT} \cite{xia2021knowledge}: This model employs a knowledge graph hierarchical transformer to capture the behavior dependencies in the recommender model.

\item \textbf{MB-STR} \cite{yuan2022multi}:  Employs multi-behavior sequential transformers to learn sequential patterns across various behaviors. 

\item \textbf{MBHT} \cite{yang2022multi}: Utilizes a hypergraph-enhanced transformer model with low-rank self-attention to model short and long behavioral dependencies. 
\item \textbf{KMCLR} \cite{xuan2023knowledge}: A knowledge graph-based approach that utilizes contrastive learning to capture the commonalities between different behaviors.
\end{itemize}


\subsection{Model performance (\textbf{RQ1})}
As shown in Tables \ref{tab:2} and \ref{tab:3}. We conducted experiments comparing HMAR to various state-of-the-art models. SASRec and SSE-PT performed well, even against multi-behavior models, due to their ability to capture sequential patterns. In contrast, MATN, MB-GCN, MB-GMN, and KHGT had weaker results, especially on the Taobao dataset. HMAR outperformed all these baselines, showing significant HR improvements on Taobao and Tianchi datasets. In comparison to the context-aware method CARCA, HMAR achieved better results with a reasonable margin on all datasets. CARCA showed competitive performance due to its use of sequential and contextual information.

\subsection{Effect of auxiliary behaviors and individual model components (\textbf{RQ2 \& RQ3})}
As shown in Table \ref{tab:7}, excluding auxiliary behaviors and relying solely on target behavior interactions has a notable impact on the Tianchi, Taobao, and Yelp datasets. This shows that auxiliary behaviors enhance the model's understanding of user behavior. Particularly, the Historical Behavior Indicator (HBI) significantly benefits the Tianchi dataset, given the influence of behaviors like page views and favorites on performance. However, there is no observable effect on the MovieLens and Yelp datasets. Multitask learning demonstrates its significance in datasets where auxiliary behaviors play a more substantial role but has a less pronounced impact on others, such as the MovieLens dataset. Additionally, eliminating the behavior encoder component, responsible for capturing deep behavioral dependencies, and relying solely on the sequence encoder, has an impact on all datasets, highlighting the advantages of the two-stage attention procedure for modeling multi-behavior sequences.

\begin{table}[h]
\centering
\scalebox{0.8}{
\begin{tabular}{c|cccc}
\hline
\backslashbox{Model}{Dataset} & Taobao & Yelp & Tianchi & MovieLens  \\
\hline
w/o Aux. behaviors   & 0.665  & 0.878  & 0.666  & 0.939 \\
w/o Multitask   & 0.846 &  0.897 &  0.781 & \textbf{0.943} \\
w/o HBI      & 0.851  &  0.901  &  0.780 & 0.942 \\
w/o Behavior Encoder      & 0.840  &  0.893  &  0.779 & 0.936 \\
\hline
HMAR   & \textbf{0.851}  & \textbf{0.901}  & \textbf{0.784}  & 0.941\\
\hline
\end{tabular}
}
\caption{Ablation study on model components (HR@10)}
\label{tab:7}
\end{table}

\section{Related Work}

\textbf{Sequential Recommendation} aims to predict the next user interaction by leveraging users' historical interactions. Early methods used Convolutional Neural Networks (CNNs) \cite{tang2018personalized} and Gated Recurrent Units (GRUs) \cite{jannach2017recurrent} for encoding interactions. SASRec \cite{kang2018self} was among the pioneers introducing the transformer architecture to sequential recommendation. Later models like BERT4Rec \cite{sun2019bert4rec} improved this approach with Bidirectional Self-Attention. Notable innovations include SSE-PT \cite{zhou2020s3} and TiSASRec \cite{li2020time}. Contemporary models like CARCA \cite{rashed2022context} incorporate contextual information and item attributes, utilizing a cross-attention mechanism for scoring. Recent approaches, like ICLRec \cite{chen2022intent}, employ contrastive techniques to boost model performance.

\textbf{Multi-behavior Recommendation} has gained traction recently due to the increased availability of diverse data sources in online applications, leading to improved recommendation accuracy. Models like MB-GCN \cite{jin2020multi} use graph convolutional networks to model multi-behavior relationships, while MB-GMN \cite{xia2021graph} employs meta-learning to address behavior heterogeneity. Recent advancements like KHGT \cite{xia2021knowledge} utilize graph attention layers for modeling relationships. Temporal aspects are considered with models like MBHT \cite{yang2022multi} and MB-STR \cite{yuan2022multi}, using attention mechanisms to incorporate sequential patterns. Cascaded graph convolutional networks \cite{cheng2023multi} further enhance understanding of behavior dependencies, and KMCLR \cite{xuan2023knowledge} introduces contrastive learning to tackle data sparsity.

In contrast to the previously mentioned models, our model combines the benefits of the two families of models, achieving superior recommendation performance.

\section{Conclusion}
In this paper, we introduced HMAR, a model designed to capture deep behavioral dependencies and sequential patterns in multi-behavioral data. It encodes items of the same behavior with a behavior encoding component, followed by a sequence encoder to capture cross-behavior dependencies. We also applied multi-task learning for behavior types. Our experiments on four multi-behavioral datasets demonstrate that HMAR outperforms state-of-the-art methods.
%
%
\bibliographystyle{splncs04}
\bibliography{sample-base}

\begin{thebibliography}{10}
\providecommand{\url}[1]{\texttt{#1}}
\providecommand{\urlprefix}{URL }
\providecommand{\doi}[1]{https://doi.org/#1}

\bibitem{bachlechner2021rezero}
Bachlechner, T., Majumder, B.P., Mao, H., Cottrell, G., McAuley, J.: Rezero is all you need: Fast convergence at large depth. In: Uncertainty in Artificial Intelligence. pp. 1352--1361. PMLR (2021)

\bibitem{chen2022intent}
Chen, Y., Liu, Z., Li, J., McAuley, J., Xiong, C.: Intent contrastive learning for sequential recommendation. In: Proceedings of the ACM Web Conference 2022. pp. 2172--2182 (2022)

\bibitem{cheng2023multi}
Cheng, Z., Han, S., Liu, F., Zhu, L., Gao, Z., Peng, Y.: Multi-behavior recommendation with cascading graph convolution networks. In: Proceedings of the ACM Web Conference 2023. pp. 1181--1189 (2023)

\bibitem{he2017neural}
He, X., Liao, L., Zhang, H., Nie, L., Hu, X., Chua, T.S.: Neural collaborative filtering. In: Proceedings of the 26th international conference on world wide web. pp. 173--182 (2017)

\bibitem{jannach2017recurrent}
Jannach, D., Ludewig, M.: When recurrent neural networks meet the neighborhood for session-based recommendation. In: Proceedings of the eleventh ACM conference on recommender systems. pp. 306--310 (2017)

\bibitem{jin2020multi}
Jin, B., Gao, C., He, X., Jin, D., Li, Y.: Multi-behavior recommendation with graph convolutional networks. In: Proceedings of the 43rd International ACM SIGIR Conference on Research and Development in Information Retrieval. pp. 659--668 (2020)

\bibitem{kang2018self}
Kang, W.C., McAuley, J.: Self-attentive sequential recommendation. In: 2018 IEEE international conference on data mining (ICDM). pp. 197--206. IEEE (2018)

\bibitem{kingma2014adam}
Kingma, D.P., Ba, J.: Adam: A method for stochastic optimization. arXiv preprint arXiv:1412.6980  (2014)

\bibitem{li2020time}
Li, J., Wang, Y., McAuley, J.: Time interval aware self-attention for sequential recommendation. In: Proceedings of the 13th international conference on web search and data mining. pp. 322--330 (2020)

\bibitem{rashed2022context}
Rashed, A., Elsayed, S., Schmidt-Thieme, L.: Context and attribute-aware sequential recommendation via cross-attention. In: Proceedings of the 16th ACM Conference on Recommender Systems. pp. 71--80 (2022)

\bibitem{sun2019bert4rec}
Sun, F., Liu, J., Wu, J., Pei, C., Lin, X., Ou, W., Jiang, P.: Bert4rec: Sequential recommendation with bidirectional encoder representations from transformer. In: Proceedings of the 28th ACM international conference on information and knowledge management. pp. 1441--1450 (2019)

\bibitem{tang2018personalized}
Tang, J., Wang, K.: Personalized top-n sequential recommendation via convolutional sequence embedding. In: Proceedings of the eleventh ACM international conference on web search and data mining. pp. 565--573 (2018)

\bibitem{wu2020sse}
Wu, L., Li, S., Hsieh, C.J., Sharpnack, J.: Sse-pt: Sequential recommendation via personalized transformer. In: Fourteenth ACM Conference on Recommender Systems. pp. 328--337 (2020)

\bibitem{xia2020multiplex}
Xia, L., Huang, C., Xu, Y., Dai, P., Zhang, B., Bo, L.: Multiplex behavioral relation learning for recommendation via memory augmented transformer network. In: Proceedings of the 43rd International ACM SIGIR Conference on Research and Development in Information Retrieval. pp. 2397--2406 (2020)

\bibitem{xia2021knowledge}
Xia, L., Huang, C., Xu, Y., Dai, P., Zhang, X., Yang, H., Pei, J., Bo, L.: Knowledge-enhanced hierarchical graph transformer network for multi-behavior recommendation. In: Proceedings of the AAAI Conference on Artificial Intelligence. vol.~35, pp. 4486--4493 (2021)

\bibitem{xia2021graph}
Xia, L., Xu, Y., Huang, C., Dai, P., Bo, L.: Graph meta network for multi-behavior recommendation. In: Proceedings of the 44th International ACM SIGIR Conference on Research and Development in Information Retrieval. pp. 757--766 (2021)

\bibitem{xuan2023knowledge}
Xuan, H., Liu, Y., Li, B., Yin, H.: Knowledge enhancement for contrastive multi-behavior recommendation. In: Proceedings of the Sixteenth ACM International Conference on Web Search and Data Mining. pp. 195--203 (2023)

\bibitem{yang2022multi}
Yang, Y., Huang, C., Xia, L., Liang, Y., Yu, Y., Li, C.: Multi-behavior hypergraph-enhanced transformer for sequential recommendation. In: Proceedings of the 28th ACM SIGKDD Conference on Knowledge Discovery and Data Mining. pp. 2263--2274 (2022)

\bibitem{yuan2022multi}
Yuan, E., Guo, W., He, Z., Guo, H., Liu, C., Tang, R.: Multi-behavior sequential transformer recommender. In: Proceedings of the 45th International ACM SIGIR Conference on Research and Development in Information Retrieval. pp. 1642--1652 (2022)

\bibitem{zhou2020s3}
Zhou, K., Wang, H., Zhao, W.X., Zhu, Y., Wang, S., Zhang, F., Wang, Z., Wen, J.R.: S3-rec: Self-supervised learning for sequential recommendation with mutual information maximization. In: Proceedings of the 29th ACM International Conference on Information \& Knowledge Management. pp. 1893--1902 (2020)

\end{thebibliography}
\end{document}